\documentclass[aps,prb,twocolumn,superscriptaddress,showpacs,showkeys,floatfix]{revtex4}

\usepackage{graphicx,color}
\usepackage{amsmath} %% wcx
\usepackage{amssymb} %% wcx

\graphicspath{{figs/}}
\begin{document}
\title{The Effects of Vacancy and Oxidation on Black Phosphorus Nanoresonators}
\author{Cui-Xia Wang$^{4}$, Chao Zhang$^{4,6}$, Jin-Wu Jiang$^{3\S}$, T. Rabczuk$^{1,2,4,5\S}$ \\
\emph{\small{$^1$Division of Computational Mechanics, Ton Duc Thang University, Ho Chi Minh City, Vietnam.}}\\
\emph{\small{$^2$Faculty of Civil Engineering, Ton Duc Thang University, Ho Chi Minh City, Vietnam.}}\\
\emph{\small{$^3$Shanghai Institute of Applied Mathematics and Mechanics, Shanghai Key Laboratory of Mechanics in Energy Engineering, Shanghai University, Shanghai 200072, People's Republic of China.}} \\
\emph{\small{$^4$Institute of Structural Mechanics, Bauhaus-University Weimar, 99423 Weimar, Germany.}}\\ 
\emph{\small{$^5$School of Civil, Environmental and Architectural Engineering, Korea University, 136-701 Korea.}} \\
\emph{\small{$^6$College of Water Resources and Architectural Engineering, Northwest A$\&$F University, 712100 Yangling, P.R. China.}} \\
\emph{\small{$^\S$Corresponding author emails: jwjiang5918@hotmail.com, timon.rabczuk@tdt.edu.vn}}}
%\date{22 December 2009}
\date{\today}
\begin{abstract}

Black phosphorene is not stable at ambient conditions, so atomic defects and oxidation effects are unavoidable in black phosphorus samples in the experiment. The effects of these defects on the performance of the black phosphorus nanoresonators are still unclear. Here, we perform classical molecular dynamics to investigate the effects of the vacancy and oxidation on single-layer black phosphorus nanoresonators at different temperatures. We find that the vacancy causes strong reduction in the quality factor of the nanoresonators, while the oxidation has weaker effect on the nanoresonators. More specifically, a 2\% concentration of randomly distributed single vacancies is able to reduce the quality factor by about 80\% and 40\% at 4.2~K and 50~K, respectively. We also find that the quality factor of the nanoresonator is not sensitive to the distribution pattern of the vacancy defects.

\end{abstract}

% 63.22.Np    Layered systems
% 63.22.-m 	Phonons or vibrational states in low-dimensional structures and nanoscale materials
\maketitle
\pagebreak

Black phosphorus (BP), a new two-dimensional nanomaterial, is composed of atomic layers of phosphorus stacked via van der Waals forces\cite{brown1965refinement}.  BP brings a number of unique properties unavailable in other two-dimensional crystal materials. For example, BP has anisotropic properties due to its puckered configuration.\cite{du2010ab,rodin2014strain,low2014tunable,engel2014black}

Currently, single-layer BP (SLBP), i.e. phosphorene,\cite{rodin2014strain} can be fabricated by mechanical exfoliation from bulk BP and has immediately received considerable attention.\cite{2014black,liu2014phosphorene,xia2014rediscovering} Atomic vacancies have been demonstrated to exist in bulk black phosphorus.\cite{zhang2009surface,baba1989preparation} Recently, first-principle calculations have demonstrated that these defects can be generated quite easily in SLBP at much higher concentrations compared with silicene and graphene.\cite{2015defects,zhang2013structures} Cai et al.\cite{cai2016highly} suggested that intrinsic itinerant behavior of atomic vacancies may result in the low chemical stability of phosphorene. In addition, another invariable issue encountered in the manipulation of SLBP is the control of the oxidation. It has been established by both theoretical calculations and experiments that $O_{2}$ can easily dissociate on black phosphorene\cite{2015oxygen,kang2015solvent} leading to the formation of the oxidized lattice.\cite{wang2015phosphorene} The presence of oxygen is suggested to be the main cause of the degradation process\cite{2015oxygen,yau1992stm} and primarily responsible for changing properties of BP\cite{huang2016interaction}, e.g., BP is turned progressively hydrophilic by oxidation. In particularly, vacancies recently have been found to have significant effects on the oxidation of phosphorene with a 5000 faster oxidizing rate at the defect site than at the perfect site.\cite{Role2016}

Very recently, Feng's group has examined the BP resonator (BPR) experimentally.\cite{wang2016resolving} Our previous work has theoretically examined the intrinsic dissipation and the effect of mechanical tension on BPRs,\cite{wang2016mechanical} in which the BPRs are perfect lattice structures without defects. However, the experimental BPR samples should have some unavoidable vacancy defects or oxidation.\cite{wang2016resolving} Furthermore, these defects can have strong effects on the performance of resonant oscillation of the nanomechanical resonators. For instance, it has been known that various defects can significantly affect the performance of the graphene nanomechanical resonators.\cite{qi2012intrinsic,zhang2014detecting,robinson2008wafer} An important task is thus to examine the effects of the vacancy defect and oxidation on the BPRs, which is the major focus of the present work.

In this work, we examine the vacancy and oxidation effects on single-layer BPR (SLBPR) via classical molecular dynamical simulations. It is found that these defects can cause a considerable degradation of the quality (Q-) factor of the SLBPRs. More specifically, a 2\% concentration of randomly distributed single vacancy (SV) is able to reduce the Q-factor by around 80\% and 40\% at 4.2~K and 50~K, respectively. The double vacancy (DV) has a slightly weaker reduction in the Q-factors of the SLBPRs. We also find that the Q-factor is mainly dependent on the percentage of the defects, while it is not sensitive to the distribution pattern of the defects. Furthermore, it is found that oxidation has a weaker reduction in the Q-factor of the SLBPRs, as compared with the vacancy defect.

\begin{figure}[tb]
  \begin{center}
    \scalebox{1.0}[1.0]{\includegraphics[width=8cm]{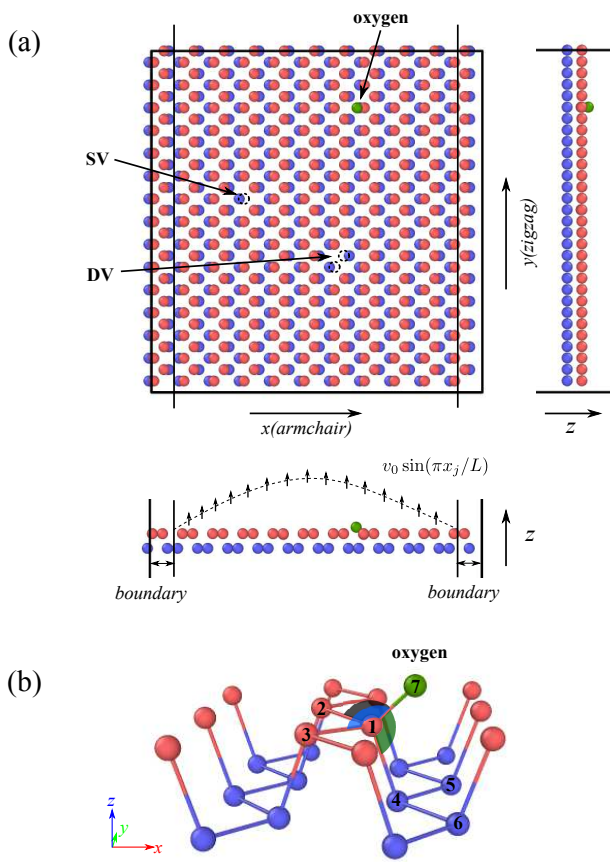}}
  \end{center}
  \caption{(Color online) (a) Configuration of SLBP with dimensions $50\times 50$~{\AA}, containing a SV defect, a DV defect and an oxidized defect. Arrows in the bottom image indicate the direction of actuation. (b) Configuration of SLBP with an oxidized defect. There are two interaction terms associated with the oxygen atom, the bond-stretching term (bond 1-7) and the angle-bending term ($\angle712$, $\angle713$ and $\angle714$). The x-axis is along the armchair direction, and the y-axis is along the zigzag direction. }
  \label{configuration}
\end{figure}

Fig.~\ref{configuration}~(a) shows the atomic structure of the SV, DV and oxidized defects in the SLBP, with a dimension of $50\times 50$ ~{\AA} that is used in our simulations. We have simultaneously considered the vacancy and oxidation effects in the present work, as density functional theory calculations have shown that these two defects usually occur together.\cite{Role2016} Vacancy defects are generated by removing atoms from the pristine SLBP. A SV defect is formed by removing one phosphorus atom, while a DV defect is formed by removing two nearest-neighboring phosphorus atoms in the same sub-layer, which has the lowest formation energy.\cite{2015defects,hu2015geometric} For the oxidized defect, the oxygen atom is put on the dangling position, which is the most stable configuration according to the previous first-principle calculations.\cite{ziletti2015oxygen,wang2015phosphorene,wang2015effects,wang2016degradation} We note that the position of the oxygen is different from the graphene oxide, where the bridge site is the preferred binding site for the oxygen atom.\cite{huang2013overcoming} As shown in Fig.~\ref{configuration}~(b), the oxygen atom bonds with one phosphorus atom, with a P-O bond (i.e., bond 1-7) length of 1.5~{\AA}, and the P-O bond is tilted by $44.5^{0}$ away from the phosphorene surface.\cite{ziletti2015oxygen,wang2016degradation} The phosphorus atom involved in the P-O bond gets dragged into the lattice structure by 0.11~{\AA} in the z direction; apart from that, the lattice deformation ignored in the present work is minimal.\cite{ziletti2015oxygen} Correspondingly, three angles $\angle OPP$ are formed, i.e., $\angle712$ ($118^{0}$), $\angle713$($118^{0}$) and $\angle714$($115^{0}$). The defect concentration is the ratio of the number of atoms removed or adsorbed over the total number of atoms in the pristine SLBP. The MD simulations are carried out using the publicly available simulations code LAMMPS,\cite{Lammps} while the OVITO package was used for visualization.\cite{ovito}

The atomic interactions among the phosphorus atoms in the structure are described by a recently-developed Stillinger-Weber potential,\cite{jiang2015parametrization} which could accurately predict the phonon spectrum and mechanical behavior of BP.\cite{jiang2015parametrization} The interaction between the oxygen atom and the SLBP is described by the bond stretching and angle bending potentials. For the bond stretching interaction, the force constant is obtained from the optimized potentials for liquid simulations (OPLS) potential, i.e., 22.776~{eV/\AA$^{2}$}.\cite{opls} The OPLS potential is able to capture essential many-body terms in interatomic interactions, including bond stretching, bond angle bending, van der Waals and electrostatic interactions.\cite{jorgensen1996development} For the angle bending interaction, there is no available value for its force constant parameter $K$. We thus treat $K$ as a simulation parameter in the following calculations; i.e., we will use different values for the force constant parameter $K$ in the MD simulations. The standard Newton equations of motion are integrated in time using the velocity Verlet algorithm with a time step of 1~$fs$.

The BPR simulations are performed as follows. First, the Nos\'e-Hoover\cite{Nose,Hoover} thermostat is applied to thermalize the entire system to a constant temperature within the NPT (i.e., the particles number N, the pressure P and the temperature T of the system are constant) ensemble for 200~{ps}. Second, the resonant oscillation of SLBP is actuated by adding a sine-shaped velocity distribution, $v_0\sin(\pi x_{j}/L)$, to the central part, as shown in the bottom panel of Fig.~\ref{configuration}~(a), while the left and right boundary parts ($2\times5$~{\AA}) are fixed. In all simulations, the velocity amplitude $v_0=2.0$~{\AA/ps} is applied, which is small enough to keep the resonant oscillation in the linear region. Third, after the actuation, the system is allowed to oscillate freely within the NVE (i.e., the particles number N, the volume V and the energy E of the system are constant) ensemble for 90~ns, and the oscillation energy is recorded to analyze the Q-factor of the SLBPRs.

\begin{figure}[tb]
  \begin{center}
    \scalebox{0.8}[0.8]{\includegraphics[width=8cm]{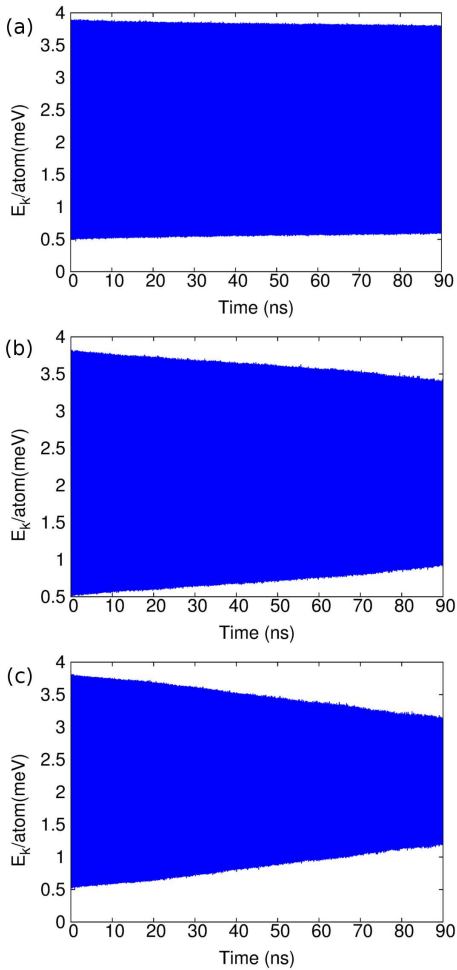}}
  \end{center}
  \caption{(Color online) The kinetic energy per atom for zigzag SLBPRs at 4.2~K with different concentrations of SV defects: 0\%, 1\% and 2\% from top to bottom. The Q-factors are 875950, 320000 and 162000 respectively.}
  \label{zig_4.2_sv}
\end{figure}
 
We first study the energy dissipation of the SLBPRs with randomly distributed SV and DV defects along the armchair and zigzag directions. Fig.~\ref{zig_4.2_sv} shows the time history of the kinetic energy per atom for the zigzag SLBPRs at 4.2~K with different concentrations of SV defects. The oscillation amplitude of the kinetic energy decays gradually, which reflects the energy dissipation during the resonant oscillation of the SLBPR. As the defect concentration increases, the energy dissipation becomes stronger, indicating a lower Q-factor with larger concentration.

\begin{figure}[tb]
  \begin{center}
    \scalebox{0.8}[0.8]{\includegraphics[width=8cm]{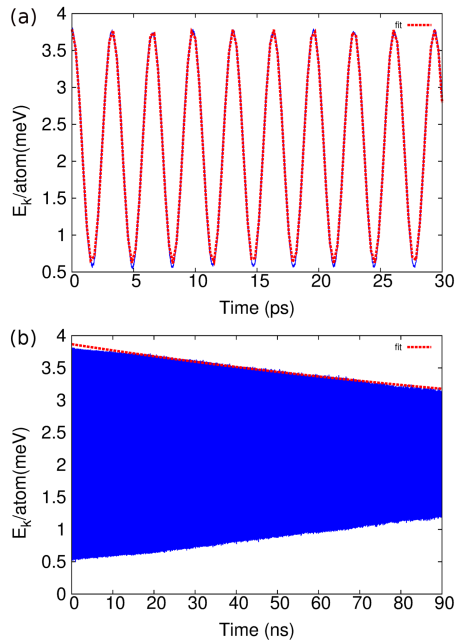}}
  \end{center}
  \caption{(Color online) Two-step fitting procedure to extract the frequency and Q-factor from the kinetic energy time history for the zigzag SLBPR at 4.2~K with 2\% SV concentration. (a) The kinetic energy time history is fitted to the function $E_{k}(t)=\bar{E}_{k}+E_{k}^0 \cos(2\pi 2ft)$ in a small time range, giving the frequency $f=0.15$~THz. (b) The kinetic energy time history is fitted to the function $E_{k}^{\rm amp}(t)=E_{k}^0 (1-\frac{2\pi}{Q})^{f t}$ in the whole time range, yielding the value of the Q-factor to be 162000.}
  \label{fit}
\end{figure}

\begin{figure}[tb]
  \begin{center}
    \scalebox{0.7}[0.7]{\includegraphics[width=8cm]{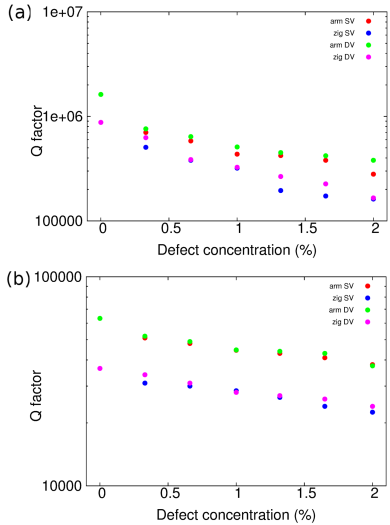}}
  \end{center}
  \caption{(Color online) Defect (SV and DV) concentration dependence for the Q-factors of the SLBPRs along the armchair and zigzag directions at (a) 4.2~K and (b) 50~K.}
  \label{sv_dv}
\end{figure}

\begin{figure}[tb]
  \begin{center}
    \scalebox{0.7}[0.7]{\includegraphics[width=8cm]{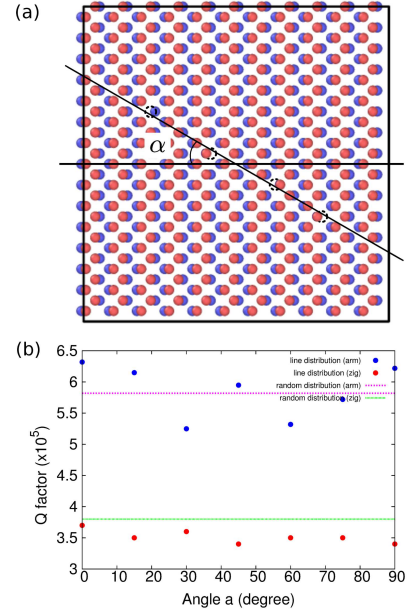}}
  \end{center}
  \caption{(Color online) The dependence of the Q-factor on the pattern of the vacancy defect. (a) The SLBP containing four SV defects (dashed circle) patterned in a line with a tilt angle $\alpha$ with respect to the oscillation direction. (b) The Q-factors of the SLBPRs along the armchair and zigzag directions at 4.2~K with different $\alpha$.}
  \label{line}
\end{figure}

\begin{figure}[tb]
  \begin{center}
    \scalebox{0.7}[0.7]{\includegraphics[width=8cm]{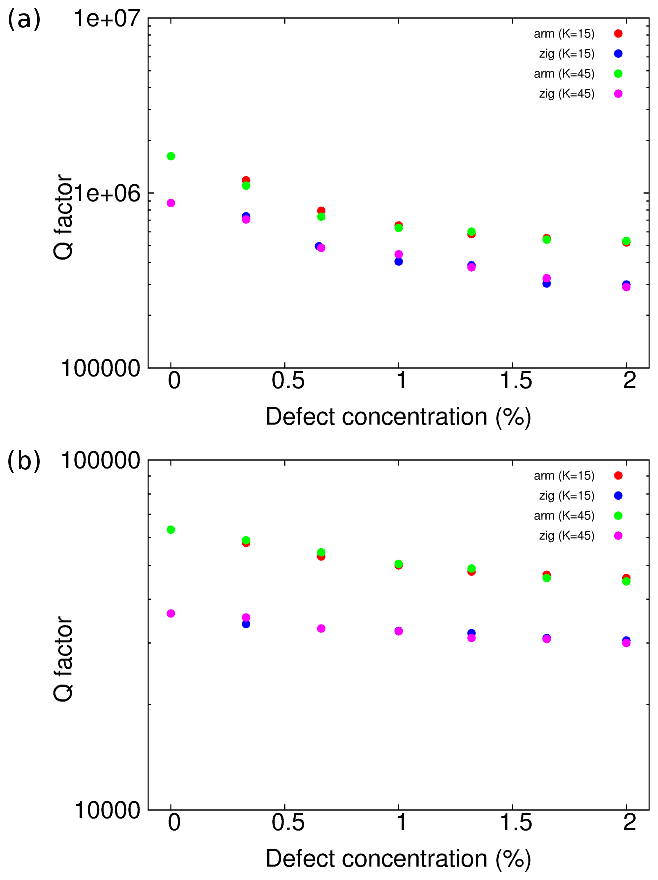}}
  \end{center}
  \caption{(Color online) Oxidized defect concentration dependence for the Q-factors of the armchair and zigzag SLBPRs at (a) 4.2~K and (b) 50~K.}
  \label{oxygen}
\end{figure}

The Q-factor and the frequency of the resonant oscillation can be extracted by fitting the kinetic energy time history to the function $E_{k}(t)=\bar{E}_{k}+E_{k}^0 \cos(2\pi 2ft) (1-\frac{2\pi}{Q})^{f t}$. The first term $\bar{E}_{k}$ represents the average kinetic energy after resonant oscillation has completely decayed. The constant $E_{k}^0$ is the total kinetic energy at $t=0$, i.e. at the moment when resonant oscillation is actuated. The frequency of resonant oscillation is $f$, so the frequency of the kinetic energy is $2f$. The energy time history is usually a very long data set, so it is almost impossible to fit it directly to the above function. The fitting procedure is thus decomposed into the following two steps as shown in Fig.~\ref{fit}. First, Fig.~\ref{fit}~(a) shows that the kinetic energy time history is fitted to the function $E_{k}(t)=\bar{E}_{k}+E_{k}^0 \cos(2\pi 2ft)$ in a very small time region $t\in [0, 30]$~ps, where the approximation $(1-\frac{2\pi}{Q})^{f t}\approx 1$ has been done for the Q-factor term as the kinetic energy dissipation is negligible in the small time range. The parameters $E_{k}^0$ and $f$ are obtained from this step. Second, Fig.~\ref{fit}~(b) shows that the oscillation amplitude of the kinetic energy can be fitted to the function $E_{k}^{\rm amp}(t)=E_{k}^0 (1-\frac{2\pi}{Q})^{f t}$ in the whole simulation range $t\in [0, 90]$~ns, which determines the Q-factor and $\bar{E}_{k}$. Following these fitting procedures, the Q-factor is 162000 for the zigzag SLBPR at 4.2~K with 2\% SV concentration. 

Fig.~\ref{sv_dv} shows the SV and DV concentration dependence for the Q-factors of the armchair and zigzag SLBPRs at 4.2~k and 50~K, respectively. With the increase of the defect concentration, the Q-factors are reduced significantly in both armchair and zigzag directions. For all concentrations considered, the Q-factor is larger in the armchair direction, which shows the same characteristic as the defect free SLBPRs.\cite{wang2016mechanical} The reduction of the Q-factor becomes weaker with the increase of the temperature. More specifically, for the SV defect with concentration of 2\%, the Q-factor drops by 82.7\% and 81.5\% in the armchair and zigzag directions, respectively, at 4.2~K, while 40\% and 38\%, respectively, at 50~K. For the DV defect, the reduction of the Q-factor is weaker. It is interesting that the vacancy effect is isotropic, i.e., the vacancy defects induce almost the same amount of reduction in the Q-factors of the armchair and zigzag SLBPRs.

The intrinsic nonlinear effect on the energy dissipation mechanism for the SLBPRs has been investigated,\cite{wang2016mechanical} in which only the phonon-phonon scattering effect is included in determining the Q-factor of the BPR. Comparing with the pristine SLBPRs, the considerable reduction of the Q-factor can be ascribed to the vacancy-induced  asymmetry of the actuated oscillation, which has been found to strongly affect the Q-factors in graphene nanomechanical resonators.\cite{qi2012intrinsic} A small percentage of defects can cause strong asymmetry in the SLBPR, leading to the reduction of the Q-factor. The defect-induced asymmetry increases with the increasing defect concentration, so the reduction of the Q-factor is also increased. As shown in Fig.~\ref{sv_dv}, the Q-factor is reduced  by 82.7\% in the armchair direction at 4.2~K with SV defect concentration of 2\%.

Experiment studies have observed that long term exposure to ambient conditions results in a layer-by-layer etching process of BP flakes. Furthermore, it is possible for flakes to be etched down to a SLBP.\cite{island2015environmental} In this etching process, the atomic vacancy is one of the unavoidable etching manner, which shall be one possible reason for low Q-factors (around 100) measured for the phosphorene resonators in the recent experiment.\cite{wang2015black} 

Recently, it has been found  that the distributed pattern of the atomic vacancies strongly affects the fracture strength and fracture strain.\cite{sha2016atomic} Thus, in addition to the randomly distributed SV defects, we have also investigated the SV defects distributed in a line pattern on the SLBP. Fig.~\ref{line}~(a) illustrates the uniformly line distribution of four SV defects, i.e. concentration of 0.33\%, where the line pattern forms an angle $\alpha$ with respect to the oscillation direction. The Q-factors of the armchair and zigzag SLBPRs at 4.2~K with different tilt angle $\alpha$ are shown in Fig.~\ref{line}~(b). It can be seen that the Q-factors of the SLBPRs show slight fluctuations with varying tilting angle $\alpha$, which means that the distribution of the vacancy defect is not important for the reduction of the Q-factor.

We now discuss the effect of oxidation on both armchair and zigzag SLBPRs. As we have noted above, the force constant parameter $K$ for the angle bending of angles $\angle712$, $\angle713$ and $\angle714$, is treated as a simulation parameter here. We have used $K=15$ and 45~{eV/rad$^{2}$} in the present work.

Fig.~\ref{oxygen} shows the defect concentration dependence for the Q-factors of the SLBPRs at 4.2~K and 50~K, respectively. It shows that the value of the parameter $K$ for the angle bending is not important here. The Q-factor is reduced significantly with the increase of the percentage. For the concentration of 2\%, the Q-factor is reduced by about 67.8\%  and 70\% in the armchair and zigzag directions, respectively, at 4.2~K, while 27.3\% and 16.3\%, respectively, at 50~K. The oxidation effect becomes weaker at higher temperature. From the above, we find that the oxidation effect is weaker than vacancy effect on the Q-factor of the SLBPRs.
 
In conclusion, we have utilized classical molecular dynamics simulations to investigate the Q-factors of the SLBPRs containing atomic vacancy or oxidized defects. It is found that these defects cause a significant degradation of the Q-factor of the SLBPRs. In particular, a 2\% concentration of randomly distributed SV is able to reduce the Q-factor by about 80\% and 40\% for the SLBPRs at 4.2~K and 50~K, respectively. Comparing with SV, the DV induces a weaker reduction in the Q-factor. We also find that the Q-factor is not sensitive to the distributed pattern of the vacancy defects. Furthermore, it is found that oxidation has a weaker reduction on the Q-factor of the SLBPRs, as compared with the vacancy defect.

\textbf{Acknowledgements} The work is supported by the China Scholarship Council (CXW and CZ). JWJ is supported by the Recruitment Program of Global Youth Experts of China, the National Natural Science Foundation of China (NSFC) under Grant No. 11504225, and the start-up funding from Shanghai University. 

\bibliographystyle{aipnum4-1}
\bibliography{biball}
%\bibliography{/home/JiangJinWu/Documents/papers/mypapers/latex/biball}

\end{document}